\documentclass[%
    reprint,
    twocolumn,
    aps,
    physrev,
    superscriptaddress,
]{revtex4-2}
\usepackage{graphicx}
\usepackage{microtype}
\usepackage[utf8]{inputenc}
\usepackage[T1]{fontenc}
\usepackage{newtxtext}
\usepackage[smallerops]{newtxmath}
\usepackage{nicefrac}
\usepackage[dvipsnames]{xcolor}
\usepackage{hyperref}
\hypersetup{
    colorlinks=true,
    linkcolor=NavyBlue,
    citecolor=NavyBlue,
    urlcolor=NavyBlue
}
\def\la{\left\langle}
\def\ra{\right\rangle}
\newcommand{\E}[1]{\times10^{#1}}
\def\del{\partial}
\def\HH{\mathcal{H}}
\def\OO{\mathcal{O}}
\def\pp{P}
\def\FF{\mathcal{F}}
\def\GG{\mathcal{G}}
\def\EE{\mathcal{E}}
\DeclareMathOperator{\Df}{D_\mathit{f}}
\DeclareMathOperator{\Dmf}{D_\mathit{f}^\mathrm{mf}}
\DeclareMathOperator{\JS}{D_{JS}}
\def\ER{Erd\H{o}s-R\'enyi}
\newcommand{\CUNY}{\affiliation{%
    Initiative for the Theoretical Sciences,
    The Graduate Center, CUNY, 
    New York, NY 10016, USA%
    }%
}
\newcommand{\UChicago}{\affiliation{%
    Department of Organismal Biology \& Anatomy
    and 
    Department of Physics,
    University of Chicago, 
    Chicago, IL 60637, USA%
    }%
}
\newcommand{\ISTAustria}{\affiliation{%
    Institute of Science and Technology Austria,
    3400 Klosterneuburg,
    Austria%
    }%
}
\newcommand{\equalcontrib}{\thanks{%
    These authors contribute equally to this work.%
    }%
}
\begin{document}\frenchspacing
\title{Inferring couplings in networks across order-disorder phase transitions}
\author{Vudtiwat Ngampruetikorn}
    \equalcontrib
    \CUNY
\author{Vedant Sachdeva}
    \equalcontrib
    \UChicago
\author{Johanna Torrence}
    \altaffiliation[Current address: ]{ShopRunner, Inc, Chicago, IL}
    \UChicago
\author{Jan Humplik}
    \ISTAustria
\author{David J.\ Schwab}
    \equalcontrib
    \CUNY
\author{Stephanie E.\ Palmer}
    \equalcontrib
    \UChicago
\begin{abstract}
Statistical inference is central to many scientific endeavours, yet how it works remains unresolved. Answering this requires a quantitative understanding of the intrinsic interplay between statistical models, inference methods and data structure. To this end, we characterise the efficacy of direct coupling analysis (DCA){\textemdash}a highly successful method for analysing amino acid sequence data{\textemdash}in inferring pairwise interactions from samples of ferromagnetic Ising models on random graphs. Our approach allows for physically motivated exploration of qualitatively distinct data regimes separated by phase transitions. We show that inference quality depends strongly on the nature of generative models: optimal accuracy occurs at an intermediate temperature where the detrimental effects from macroscopic order and thermal noise are minimal. Importantly our results indicate that DCA does not always outperform its local-statistics-based predecessors; while DCA excels at low temperatures, it becomes inferior to simple correlation thresholding at virtually all temperatures when data are limited. Our findings offer new insights into the regime in which DCA operates so successfully and more broadly how inference interacts with data structure.
\end{abstract}
\maketitle
\section*{Introduction}
A quantitative understanding of the limitations and biases of inference methods is critical for developing high performing and trustworthy approaches to data analyses. While emerging, such an understanding is incomplete, not least because it requires a thorough investigation of the intertwined nature of statistical models, inference methods and data structure~\cite{Zdeborova:20}. Statistical physics models are ideally suited for this investigation for three main reasons. First, they often encompass the statistical models used in practice; take, for example, the Potts model in direct coupling analysis (DCA)~\cite{Weigt:09,Morcos:11}. Second, they enjoy a number of well-studied inference methods owing to a long history of inverse statistical physics problems~\cite{Roudi:09,Nguyen:17,Cocco:18}. Third, they provide a controlled and physically motivated way to alter data structure across qualitatively distinct regimes. Adopting a statistical physics approach, we characterise the performance of DCA, one of the most oft-used tools in biological sequence analyses, and highlight the importance of data structure in quantifying the performance of inference methods.

DCA has proved successful as a technique for inferring the physical interactions that underpin the structure of biological molecules from amino acid sequence data~\cite{Weigt:09,Morcos:11}. This success has led to new insights into the protein folding problem~\cite{Marks:11} and how RNAs obtain their structures~\cite{DeLeonardis:15,Weinreb:16,Wang:17}. The essence of DCA is to draw a distinction between direct and indirect correlations{\textemdash}those originating from direct physical interactions between two sites in a sequence and those mediated via other sites{\textemdash}by fitting a global statistical model to sequence data. But while DCA supersedes its local-statistics-based predecessors in virtually all applications, relatively little is known about the conditions that underlie its success~\cite{Kleeorin:21}.

\begin{figure}
\includegraphics[width=\linewidth]{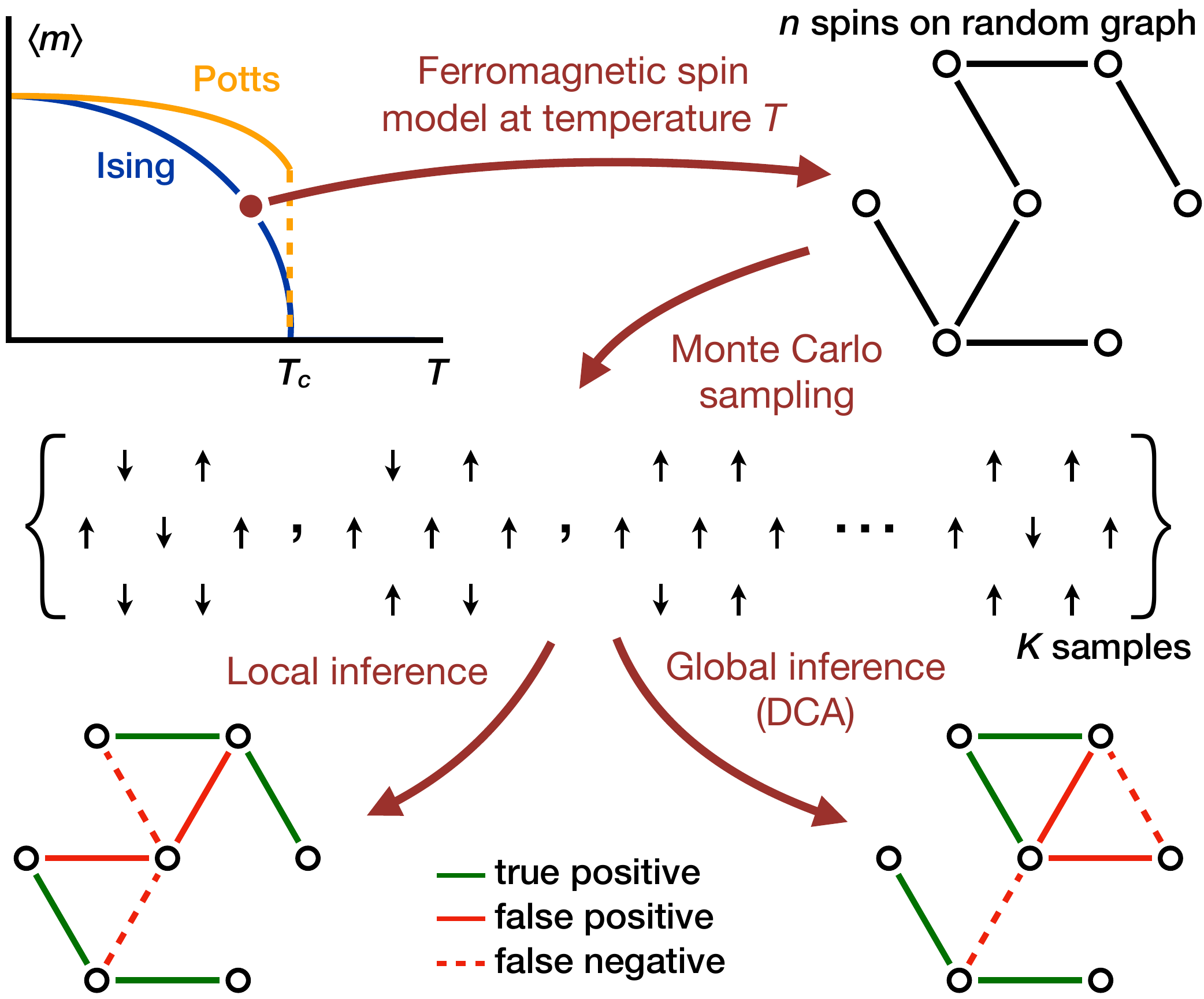}
\caption{%
\label{fig:schematic}%
\textbf{Data generation and inference.}
We generate samples from a ferromagnetic spin model on an \ER\ random graph and evaluate inference methods on the data at different model temperatures across order-disorder phase transitions. Direct coupling analysis ranks the likelihood of an interaction by leveraging global statistics whereas local inference uses pairwise statistics such as empirical correlations. We obtain predictions by thresholding the likelihood scores.%
}
\end{figure}

The statistical model in DCA, well-known in physics as the Potts model~\cite{Wu:82}, captures a phase transition that results from a competition between disorder-promoting thermal noise and order-promoting interactions. The disordered phase, which prevails at high temperatures, describes a system whose constituents (e.g., residues in a sequence) are largely uncorrelated; on the other hand, a macroscopic number of such constituents assume the same state in the low-temperature ordered phase. Both phases make for difficult inference: the data are noisy in the disordered phase and macroscopic ordering leads to strong indirect correlations in the ordered phase~\cite{Montanari:09}. A question arises as to the regime in which DCA operates so successfully and more broadly how the nature of data-generating distributions affects inference (see, also Ref~\cite{Bialek:20}).

Recent work suggests that sequence data are drawn from distributions poised at the onset of order~\cite{Tang:17,Tang:20}. This regime sits at the boundary of the two phases, thus minimising the detrimental effects from thermal noise while avoiding precipitation of macroscopic order. In fact signatures of criticality{\textemdash}a defining property of a type of phase transitions{\textemdash}appear ubiquitous across a wide variety of biological systems~\cite{Mora:11,Bialek:17}, including antibody diversity~\cite{Mora:10}, genetic regulations~\cite{Nykter:08,Krotov:14}, neural networks~\cite{Levina:07,Chialvo:10,Tkacik:13,Tkacik:15,Mora:15,Chen:19,Meshulam:19}, behaviours of individuals~\cite{Costa:19} and those of groups~\cite{Bialek:14,Attanasi:14a,Attanasi:14b}. This apparent ubiquity has inspired a search for the origin of this behaviour~\cite{Schwab:14,Aitchison:16,Morrell:20} as well as work that attempts to uncover its function~\cite{Meijers:21}. However the structure of data distributions alone cannot capture the complete phenomenology of inference and as such cannot explain the success of DCA relative to local-statistics-based methods.

Here we investigate the efficacy of DCA in inferring pairwise couplings from samples drawn from ferromagnetic spin models on random graphs at different temperatures across order-disorder phase transitions, see Fig~\ref{fig:schematic}. We demonstrate that the inference quality depends on data structure; in particular, better inference methods need not be more elaborate nor computationally more expensive. We show that a simple method based on thresholding pairwise correlations can easily outperform DCA at all temperatures when data are sparse{\textemdash}a condition applicable to nearly all amino acid sequence datasets. We find further that more data improve DCA most significantly in the ordered phase where strong indirect correlations limit the performance of local methods. Interestingly we do not observe direct effects of criticality despite its association with diverging Fisher information~\cite{Brody:95,Janke:04,Crooks:07,Mastromatteo:11,Prokopenko:11}. Instead we attribute the accuracy maximum at an intermediate temperature to the competition between the emergence of macroscopic order at low temperatures and high thermal noise level at high temperatures. Our work underscores the necessity to characterise the role of data structure when evaluating inference methods and offers a first step towards a deeper understanding of the intertwined nature of inference, models and data structure.

\section*{Results}

To highlight the role of a phase transition, we consider the problem of reconstructing the interaction matrix of an Ising model on a random graph. 
A limiting case of the Potts model, the Ising model is one of the simplest models that captures a phase transition. 
It describes a system of $n$ spins, $\vec{\sigma}\!=\!(\sigma_1,\sigma_2,\dots,\sigma_n)$, each of which is a binary variable $\sigma_i\!\in\!\{\pm1\}$. 
The spins interact via the Hamiltonian 
\begin{equation}
\label{eq:H_Ising}
\HH(\vec{\sigma}) = -\sum_{i=1}^n\sum_{j=i+1}^n J_{ij}\sigma_i\sigma_j - \sum_{i=1}^nh_i\sigma_i,
\end{equation}
where $J_{ij}$ denotes the interaction between spins $i$ and $j$, and $h_i$ the bias field on spin $i$. 
The probability distribution of this system is given by
\begin{equation}\label{eq:prob}
\pp(\vec \sigma) = \frac{e^{-\beta\HH(\vec \sigma)}}{\sum_{\vec\sigma'} e^{-\beta\HH(\vec \sigma')}},
\end{equation}
where $\beta\!=\!1/T$ is the inverse temperature and the summation is over all spin configurations. 

Fig~\ref{fig:schematic} provides an overview of our work. We generate samples from a uniform-interaction ferromagnetic Ising model on an \ER\ random graph,
\begin{equation}
\label{eq:H_0}
\HH^\text{data}(\vec{\sigma}) 
=
-\sum\nolimits_{i<j}J_{ij}\sigma_i\sigma_j
\quad\text{with}\quad
J_{ij}\sim \mathrm{Bern}(\lambda/n)
\end{equation}
for a graph with $n$ vertices and mean degree $\lambda$. 
Each interaction is drawn from a Bernoulli distribution with parameter $p\!=\!\lambda/n$, i.e., an interaction is present ($J_{ij}\!=\!1$) with probability $p$ and absent ($J_{ij}\!=\!0$) with probability $1-p$. 
In the thermodynamic limit $n\!\to\!\infty$, a sharp transition exists between the high-temperature disordered phase and the low-temperature ordered phase. 
This phase transition is characterised by the order parameter $\Delta\!\equiv\!\frac{1}{n}|\langle\sum_i\sigma_i\rangle|$, which vanishes in the disordered phase and grows continuously with decreasing temperature in the ordered phase. 
A standard mean-field approximation yields the critical temperature $T_c\!=\!\lambda$ with the order parameter given by the largest root of the equation $\Delta\!=\!\tanh(\lambda\Delta/T)$.
Our results are based on samples generated with exact Monte Carlo sampling~\cite{Propp:96}.

While several methods exist for the inverse Ising problem~\cite{Nguyen:17}, we focus on the so-called naive mean-field inversion which forms the basis for a number of practically relevant algorithms~\cite{Roudi:09,Morcos:11,Marks:11,Stein:15}. 
Derived from a mean-field theory and the linear response theorem~\cite{Kappen:98,Tanaka:98}, the naive mean-field inversion expresses interactions $J_{ij}$ in terms of empirically accessible connected correlation matrix $C$, 
\begin{equation}\label{eq:nMF_inversion}
\beta J_{ij} = -(C^{-1})_{ij}
\quad\text{for $i<j$},
\end{equation}
where 
$C_{ij}\!\equiv\!\langle\sigma_i\sigma_j\rangle-\langle\sigma_i\rangle\langle\sigma_j\rangle$.
In the following, global statistical inference refers to the naive mean-field inversion.

\begin{figure*}
\centering
\includegraphics[width=\linewidth]{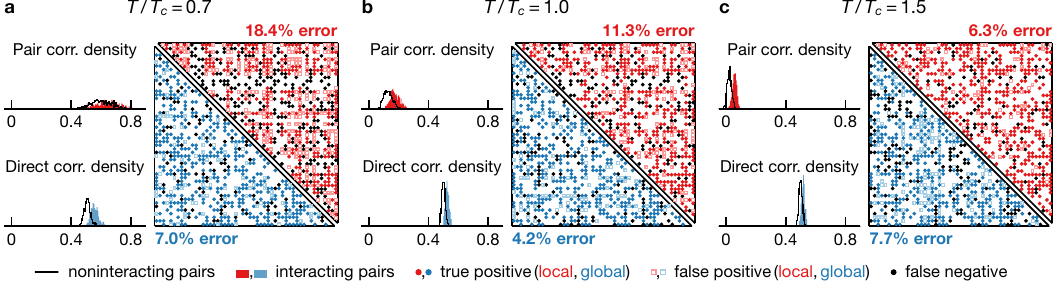}
\caption{\label{fig:contactmap}%
\textbf{%
Local statistical modelling outperforms mean-field DCA in the disordered phase.}
We show density histograms of empirical and direct pair correlations{\textemdash}$\langle\sigma_i\sigma_j\rangle_\text{data}$ and $\langle\sigma_i\sigma_j\rangle_\text{dir}$ [see, Eq~\eqref{eq:p_direct}]{\textemdash}for interacting (filled) and non-interacting (line) pairs of spins at $T/T_c\!=\!0.7,1.0,1.5$ (\textbf{a-c}, respectively). 
The predictions of pairwise interactions are depicted in a contact map for local (upper half) and global (lower half) inference. 
The discrimination threshold is chosen such that the number of positive predictions is equal to the number of real interactions, and false positives and false negatives are equal (see legend). 
In general both empirical and direct pair correlations are higher among interacting spins and are thus informative of interactions. 
For local inference, the prediction error decreases with temperature and is smaller than that of global inference at $T/T_c\!=\!1.5$ (\textbf{c}).
Global inference error exhibits non-monotonic temperature dependence and is minimal at an intermediate temperature $T/T_c\!=\!1.0$ (\textbf{b}). 
Shown results are based on $5\E3$ samples drawn from an Ising model on an \ER\ graph with $50$ vertices and mean degree $20$.
}
\end{figure*}

\begin{figure}
\centering
\includegraphics{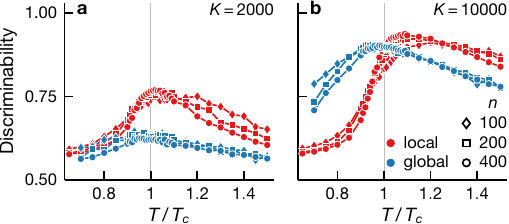}
\caption{\label{fig:discriminability}%
\textbf{%
Local inference is more data efficient but more severely affected by macroscopic order.%
}
We depict the local (red) and global (blue) inference discriminability of interactions (area under the ROC curve) for Ising models on \ER\ graphs with mean degree 40 and different number of vertices $n$ (see legend) for sample sizes $K\!=\!2\E3$ and $10^4$ (\textbf{a} and \textbf{b}, respectively). 
Both local and global inference exhibits discriminability maximum near $T_c$. 
Local inference is more discriminating at all temperatures when the data are limited (\textbf{a}). 
But global inference performs better in the ordered phase when more data are available (\textbf{b}). 
}
\end{figure}

One measure of inference quality is the ability to discriminate directly interacting spin pairs from those that interact only via other spins. 
Fig~\ref{fig:contactmap} visualises this discrimination based on local and global statistical inference. For each spin pair, we assign a score that ranks the likelihood of an interaction being present; here, we use empirical correlations $\langle\sigma_i\sigma_j\rangle_\text{data}$ and direct correlations $\langle\sigma_i\sigma_j\rangle_\text{dir}$ in local and global inference, respectively. The average $\langle\cdots\rangle_\text{dir}$ is taken with respect to the direct pairwise distribution~\cite{Weigt:09}, 
\begin{equation}\label{eq:p_direct}
\hat\pp^\text{dir}_{ij}(\sigma_i,\sigma_j)\equiv
\frac{
\exp(\beta\hat{J}_{ij}\sigma_i\sigma_j+\tilde h_i\sigma_i+\tilde h_j\sigma_j)
}{
\sum_{\sigma'_i,\sigma'_j}\exp(\beta\hat{J}_{ij}\sigma'_i\sigma'_j+\tilde h_i\sigma'_i+\tilde h_j\sigma'_j)
}
\end{equation}
where $\hat{J}_{ij}$ denotes the inferred interactions from naive mean-field inversion and the fields $\tilde h_i$ and $\tilde h_j$ are chosen such that the marginal distributions coincide with empirical single-spin distributions.
In Fig~\ref{fig:contactmap}, we see that on average both empirical and direct correlations are higher among interacting pairs and are thus predictive of true interactions. 
To turn the likelihood scores into concrete predictions, we need to define a threshold which separates positive and negative predictions. 
We choose a discrimination threshold that equates the number of positive predictions to the number of true interactions and display inference predictions and errors as a contact map (Fig~\ref{fig:contactmap}a-c). 
The accuracy of the global approach exhibits non-monotonic temperature dependence with higher error rates at temperatures above and below $T_c$.
In contrast the accuracy of local inference increases with temperature over the range shown in Fig~\ref{fig:contactmap}.
(But note that the accuracy must eventually go down at adequately high temperatures, see Fig~\ref{fig:discriminability}.)
While the error rate of global inference is less than half of that of local inference at low temperatures (Fig~\ref{fig:contactmap}a-b), a local statistical approach outperforms global inference at high temperature (Fig~\ref{fig:contactmap}c; see also, Fig~\ref{fig:discriminability}). 

Although specifying a discrimination threshold allows us to make concrete predictions, its choice is often arbitrary. 
We now consider a more general measure of discriminability grounded in receiver operating characteristic (ROC) analysis. 
ROC analysis constructs a curve that traces the true and false positive rates as the discrimination threshold varies.  
In the following, we identify discriminability with the area under the ROC curve which is equal to the probability that a real positive scores higher than a real negative.

Local and global statistical inference exhibits qualitatively different sample size dependence, see Fig~\ref{fig:discriminability}. 
At low samples, local inference is more discriminating than naive mean-field inversion at all temperatures (Fig~\ref{fig:discriminability}a). 
This behaviour is a result of the distinct natures of local and global approaches. 
Global inference requires a good estimate of the full joint distribution whereas local inference relies only on pairwise distributions which are much easier to estimate, especially with limited samples. 
An increase in samples improves both local and global inference but this improvement diminishes for local inference at low temperatures (Fig~\ref{fig:discriminability}b).
This results from the fact that the entropy of the model increases with temperature and thus, given a fixed number of samples, a low-temperature model is better sampled. 
In Fig~\ref{fig:discriminability}a, pairwise distributions are already well-sampled at low temperatures and more samples do not lead to higher accuracy for local inference (Fig~\ref{fig:discriminability}b). 
However well-sampled pairwise distributions do not imply a good estimate of the full distribution; indeed, we see that the discriminability of global inference goes up with samples at low temperatures (Fig~\ref{fig:discriminability}a-b). 
%
\begin{figure*}
\centering
\includegraphics{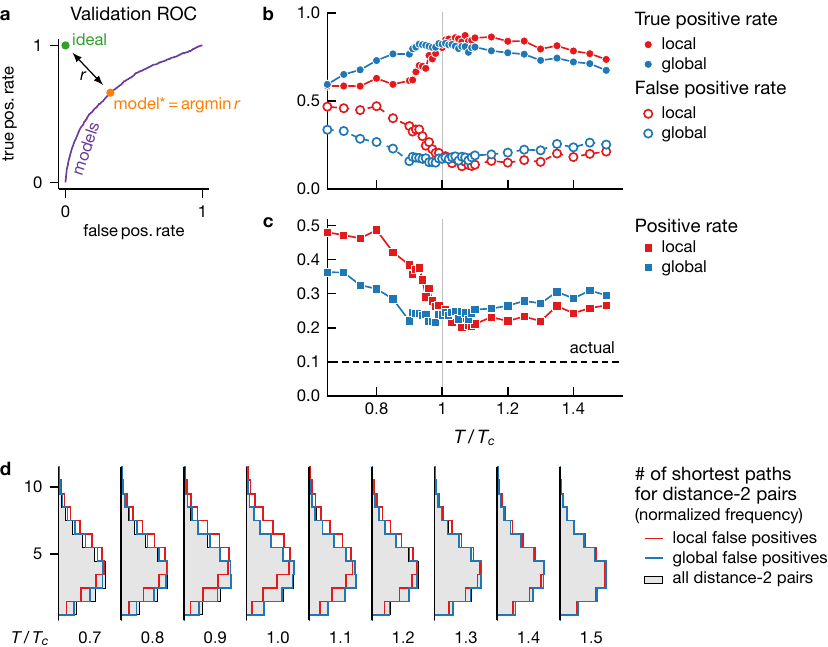}
\caption{\label{fig:validatedmodel}%
\textbf{%
Local inference is more likely to mis-classify well-connected non-interacting pairs.%
}
We use 20\% of pairs chosen at random (validation set) to compute the discrimination threshold (\textbf{a}) and report inference properties on the rest (test set, \textbf{b-d}). 
\textbf{a} Typical ROC curve for the validation set. 
We choose a threshold such that the resulting model is closest to the ideal model, as measured by the Euclidean distance in the ROC space. 
\textbf{b}
True and false positive rates \textit{vs} temperature.
Both local and global methods are most accurate at a temperature close to $T_c$ but local inference worsens faster at low temperatures. 
\textbf{c}
Temperature dependence of the positive rate (the ratio between positive predictions and all pairs). 
Over-prediction is most acute for local inference at low temperatures. 
\textbf{d}
Distribution of the number of shortest paths among false positive pairs with graph distance two at different temperatures. 
At low temperatures the false positives from local inference contain a larger fraction of highly connected pairs, compared to all pairs with distance two (grey) as well as to the false positives from global inference. 
Thus non-interacting pairs in denser parts of the graph are likelier to be mis-classified than those in sparser parts. 
Shown results are based on $10^4$ samples from an Ising model on an \ER\ graph with 400 vertices and mean degree 40. 
}
\end{figure*}

Inference performance depends not only on well-measured probability distributions but also the structure of the distributions. 
Despite having lower entropy and being better sampled, low-temperature models are more difficult to infer compared to those in the vicinity of the phase transition, see Fig~\ref{fig:discriminability}. 
This feature is a consequence of macroscopic ordering below $T_c$. 
In the ordered phase, two spins are likely to align regardless of the presence of an interaction and therefore pair correlations become less discriminating. 
While the decrease in discriminability affects both local and global inference, its effect is less severe for global inference (Fig~\ref{fig:discriminability}). 
The use of global statistics helps avoid direct comparisons between spin pairs in dense clusters of the interaction graph and those in sparser parts.

Indeed local inference is more likely to mis-classify well-connected non-interacting spin pairs. 
To illustrate this point, we randomly divide the pairs into two disjoint sets for validation and testing. 
We use the validation set to determine a discrimination threshold and report inference quality on the test set. 
In Fig~\ref{fig:validatedmodel} we use 20 percent of pairs in validation and choose the discrimination threshold such that the resulting true and false positive rates are closest to that of ideal classifiers, as measured by the Euclidean distance in the ROC plane (Panel A). 
Fig~\ref{fig:validatedmodel}b and c show that the quality of local inference deteriorates faster as temperature decreases below $T_c$---i.e., decreasing true positive rate, increasing false positive rate and more over-prediction. 
We characterise the false positives (mis-classified non-interacting pairs) by the number of shortest paths between spins in each pair (Fig~\ref{fig:validatedmodel}d). 
Here we focus only on pairs with a graph distance of two (less than two percent of pairs have distance greater than two for this particular graph). 
At high temperatures the distribution of the number of shortest paths among false positives is the same as that for non-interacting pairs; that is, any non-interacting pair is equally likely to be mis-classified. 
As temperature lowers to around $T_c$, the false positives from local inference contain a disproportionately large fraction of pairs that are connected by more paths. 
This behaviour is a direct consequence of the emergence of order which generates strong correlations, especially among pairs in denser parts of the graph. 
At very low temperatures, macroscopic order proliferates and pair correlations are strong regardless of the number of paths or physical interactions. 
While this effect reduces the disproportionate mis-classification among better connected pairs, it increases the discrepancy between the predicted and actual positive rates (Fig~\ref{fig:validatedmodel}c).
In fact the positive rate of $\sim\!\!50$ percent results from the fact that any pair leads to a positive prediction with probability $\nicefrac{1}{2}$. 
We see that in contrast to local inference, mean-field DCA is less likely to confound path multiplicity with interactions, especially close to the onset of order. 
In addition it suffers less from strong indirect correlations as evidenced by smaller over-prediction rates at low temperatures.
In sum, leveraging global statistics helps DCA draw a better distinction between direct and indirect correlations, thus making it more accurate at low temperatures.

\begin{figure}
\centering
\includegraphics{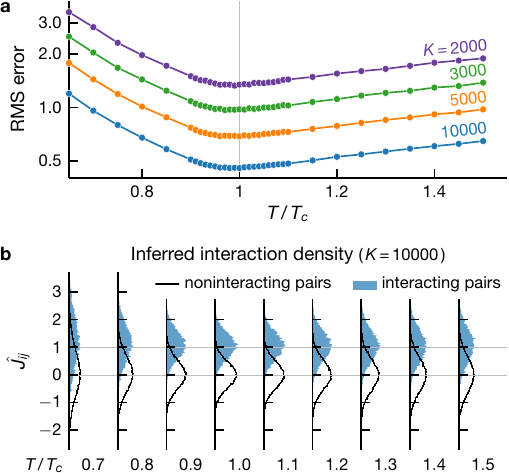}
\caption{\label{fig:rms}%
\textbf{%
Interactions inferred from mean-field DCA are statistically unbiased with smallest variances around phase transitions. 
}
\textbf{a} Root-mean-square error of inferred interactions as a function of temperature at different sample sizes $K$ (see legend). 
\textbf{b} Density histograms of inferred interactions for non-interacting and interacting pairs whose true interactions are one and zero, respectively. 
Shown results are for an Ising model on an \ER\ graph with 400 vertices and mean degree 40. 
}
\end{figure}

While a useful characterisation of discriminability, ROC analysis is agnostic about the magnitude of the inferred interactions. 
We now show that the root-mean-square (RMS) error of the interactions inferred by naive mean-field inversion exhibits similar temperature dependence to discriminability. 
In Fig~\ref{fig:rms}a, we see that the RMS error is smallest at a temperature slightly below $T_c$ for a range of sample sizes.
Fig~\ref{fig:rms}b reveals the origin of this temperature dependence. 
On average mean-field inversion correctly predicts the interactions---$J_{ij}\!\in\!\{0,1\}$ depending on whether an interaction is present---but the prediction variance is minimum around $T_c$. 
Above $T_c$, an increase in temperature leads to a model with higher entropy, thus requiring a larger number of samples to maintain inference accuracy. 
Below $T_c$, macroscopic order interferes with inference by generating strong indirect correlations among non-interacting pairs. 

Since inference quality is intrinsically a combined property of inference methods and data distributions, it is \emph{a priori} unclear whether the observed non-monotonic temperature dependence (Figs~\ref{fig:discriminability} and \ref{fig:rms}) originates from the inductive bias in inference methods or the data structure. 
To isolate the role of data-generating models, we consider the response of data distributions to a change in model parameters as a proxy for how informative a data point is about model parameters. 
We quantify the distributional response by the \textit{f}-divergence, an information-theoretic distance between two distributions, defined via $\Df(P_X\|Q_X)\!\equiv\!\langle f(P_X/Q_X)\rangle_{X\sim Q_X}$ where $f$ is convex and $f(1)\!=\!0$.
The \textit{f}-divergence between two zero-field Ising models on different graphs, parametrised by $J$ and $J'$, reads [see, Eqs~\eqref{eq:prob} and \eqref{eq:H_0}]
\begin{equation}\label{eq:Df}
\Df(J'\!,J)
=
\la
f
\left(\frac{
e^{\beta\sum_{i<j}\Delta J_{ij}\sigma_i\sigma_j}
}{
\la
e^{\beta\sum_{i<j}\Delta J_{ij}\sigma'_i\sigma'_j}
\ra_{\vec\sigma'\sim\HH_J}
}
\right)
\ra_{\!\!\vec\sigma\sim\HH_J}\!,
\end{equation}
where $\Delta J\!=\!J'-J$ and the average $\langle\dots\rangle$ is with respect to the model on the graph $J$. 

Before we discuss the numerical results, it is instructive to derive an expression for the \textit{f}-divergence in a mean-field approximation. 
Expanding Eq~\eqref{eq:Df} around $\beta\!=\!0$ and taking $P(\vec\sigma)\!=\!\prod_i \tfrac{1}{2}(1+\sigma_i\Delta)$ yield
\begin{equation}\label{eq:Dmf}
    \Dmf(J'\!,J)=
    \frac{1}{2}f''(1)\|\Delta{J}\|_1\frac{1-\Delta(T)^4}{T^2},
\end{equation}
where $\Delta(T)$ is the mean-field order parameter and the $\ell_1$-norm $\|\Delta{J}\|_1$ counts the number of different edges in $J$ and $J'$. 
Note that the elements of $J$ and $J'$ are either zero or one and we set $J_{ij}\!=\!0$ for $i\!\ge\!j$ as they do not enter the model [see, Eq~\eqref{eq:H_0}]. 
In the disorder phase $T\!>\!T_c$, high noise level makes models less dependent on the parameters and the \textit{f}-divergence decays as $T^{-2}$. 
The dependence on the order parameter means different parameters also result in more similar models at low temperatures. 
Indeed the competition between thermal noise and macroscopic order leads to a maximum at $T/T_c\!\approx\!0.83$. 

\begin{figure}
\centering
\includegraphics{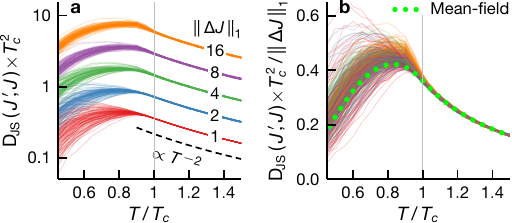}
\caption{\label{fig:js}%
\textbf{%
Jensen-Shannon (JS) divergence between two Ising models \textit{vs}
temperature.}
\textbf{a} JS divergences computed from $10^4$ samples using Eq~\eqref{eq:Df} for a fixed graph $J$ and many realisations of $J'$ generated by randomly deleting and adding edges to $J$. 
The curves are grouped by the number of different edges in $J$ and $J'$ (see legend). 
\textbf{b} empirical JS divergences compared to a mean-field prediction Eq~\eqref{eq:Dmf}, showing good agreement for $T\!>\!T_c$ (same colour code as in \textbf{a}). 
Here $J$ is an \ER\ graph with $400$ vertices and mean degree $40$.  
}
\end{figure}
Fig~\ref{fig:js} illustrates the temperature dependence of the \textit{f}-divergence between two Ising models. 
Here we adopt the Jensen-Shannon (JS) divergence which is an \textit{f}-divergence defined with $f(t)\!=\!(t+1)\log_2\frac{2}{t+1}+t\log_2 t$. 
We compute the divergence $\JS(J'\!,J)$ from data using Eq~\eqref{eq:Df} for a fixed \ER\ graph $J$ and we generate $J'$ by randomly deleting and adding edges in $J$.
We see that, as expected from the mean-field analysis, the \textit{f}-divergence decays as $T^{-2}$ at high temperatures and peaks at a temperature below $T_c$ with its scale controlled by the number of different interactions in $J$ and $J'$ (Fig~\ref{fig:js}a). 
In Fig~\ref{fig:js}b, we compare the empirical JS-divergence to the mean-field approximation [Eq~\eqref{eq:Dmf}] and find good agreement for $T\!>\!T_c$. 
Below $T_c$, the mean-field result only captures the qualitative behaviour due to large variance in the JS divergence (from different realisations of $J'$). 
This is an expected result since the locations where macroscopic order nucleates depend on graph structure and a change to which can yield a range of divergences.

It is tempting to view the inference quality maximum as a manifestation of critical phenomena, not least because the Fisher information (magnetic susceptibility) diverges at $T_c$~\cite{Brody:95,Janke:04,Crooks:07,Mastromatteo:11,Prokopenko:11}. 
However criticality does not seem to play an important role in inferring the interaction graph. 
Indeed Fig~\ref{fig:js} illustrates that the distance between two models on different graph varies smoothly across the critical temperature.

\begin{figure}
\centering
\includegraphics{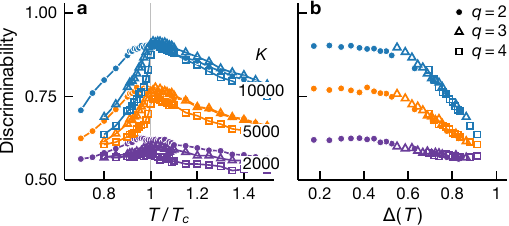}
\caption{\label{fig:potts}%
\textbf{%
Interaction discriminability for Ising and Potts models.
}%
Discriminability maximum results from the competition between thermal noise and macroscopic ordering but is not a signature of criticality associated with second-order phase transitions. 
We show DCA discriminability at different sample sizes $K$ (see legend) as a function of temperature (\textbf{a}) and mean-field order parameter $\Delta$ (\textbf{b}). 
In all cases, discriminability peaks at an intermediate temperature and displays similar temperature dependence above $T_c$. 
By plotting discriminability as a function of $\Delta$ for $T\!<\!T_c$, we see that different temperature dependence for Ising and Potts models at $T\!<\!T_c$ originates from the fact that macroscopic order forms more rapidly in Potts models which admit first-order phase transitions. 
This highlights the detrimental effect of macroscopic order on inference quality.  
Shown results are based on the same \ER\ interaction graph with 400 vertices and mean degree 40. 
}
\end{figure}

To elaborate this point further, we consider $q$-state Potts models on an \ER\ random graph which generalises the binary spins in Ising models to $q$ states. 
Unlike the Ising model, a $q$-state Potts model with $q\!>\!2$ exhibits a discontinuous phase transition which does not display critical behaviours and at which the susceptibility remains finite. 
Fig~\ref{fig:potts} compares the inference discriminability for 3 and 4-state Potts models with that for Ising models ($q\!=\!2$). 
We use the naive mean-field inversion, generalised to Potts models~\cite{Morcos:11} for both Ising and Potts models (see, Mean-field inversion in Methods).
In Fig~\ref{fig:potts}, we see that, in the disordered phase, the discriminability for Potts and Ising models shows similar dependence on sample size and temperature.
In the ordered phase, the inference quality decreases with temperature and worsens with increasing $q$. 
This $q$-dependence results from the fact that macroscopic order forms more rapidly for larger $q$ with order parameter discontinuity growing with $q$, see Phase transitions in Potts models on homogeneous random graphs in Methods [Eq \eqref{eq:potts_discontinuity}].
In fact, Fig~\ref{fig:potts}b illustrates that the inference discriminability for Potts and Ising cases displays similar dependence on the mean-field order parameter (for a mean-field analysis of the Potts model, see Ref~\cite{Wu:82} and Graphical Potts models in Methods), thus suggesting that macroscopic ordering rather than criticality is an important determinant of inference performance. 

\section*{Discussion}
Despite being more elaborate and computationally more expensive than local statistical approaches, mean-field DCA does not always lead to better inference quality. Indeed we show that local statistical methods can be more accurate when data are limited. More generally, although global statistics encode more information that could potentially improve inference, they are more difficult to estimate from sparse data. Inference quality depends not only on well-measured statistics but also the nature of data distributions. A low-temperature generative model, while better-sampled due to lower entropy, is more difficult to infer, compared to higher-temperature models around the phase transition. This feature highlights how macroscopic ordering, and more broadly data structure, can interfere with inference. For models exhibiting an order-disorder phase transition, we find that DCA provides the most advantage over local statistical modelling in the ordered phase and when the systems are relatively well-sampled. Our results highlight the fact that inference quality can only be quantified with respect to data structure and illustrate the central role of data structure in understanding inductive biases of inference methods~\cite{Das:20}. Finally our work lays a foundation for future investigations seeking to provide a prescription for inference method selections based on data structure.

The use of the Potts model to capture correlations among constituents of a system is neither unique to DCA nor limited to analysing sequence data. Indeed this approach is applicable to a range of biological systems from neural activity~\cite{Schneidman:06,Tkacik:06} to flocks of birds~\cite{Bialek:12}. In particular our findings on the efficacy of local statistical inference are directly relevant to neural and other systems for which data tend to be sparse (see also Ref~\cite{Kleeorin:21}). In addition, the Potts model is closely related to probabilistic graphical models and Markov random fields in probability theory, statistics and machine learning with applications including inferring interactions among genetic transcription factors~\cite{Friedman:04} and computer vision~\cite{Wang13}. Our results thus have implications for a large class of problems beyond the application of DCA in structural biology.

To isolate the role of a phase transition, we specialise our analysis to uniform-interaction models on \ER\ random graphs which tend to be less structured than interaction graphs of real systems. For example, the structural organisation of proteins leads to a hierarchy of sectors of strongly interacting amino acids~\cite{Halabi:09}. Spin models on hierarchical random graphs also capture order-disorder phase transitions~\cite{Dorogovtsev:02} and it would be interesting to investigate how such a structure affects inference. Another promising future direction is to extend our analysis beyond ferromagnetic models to systems with richer phase diagrams such as spin-glass models and sparse Hopfield networks. 

\section*{Methods}
\subsection*{Graphical Potts models \label{GPM}}
Potts models describe a system of $q$-state spins $\vec{\sigma}\!=\!(\sigma_1,\sigma_2,\dots,\sigma_n)$ with $\sigma_i\!\in\!\{1,2,\dots,q\}$, interacting via the Hamiltonian,
\begin{equation}
\label{eq:HH}
\HH(\vec{\sigma}) = -\sum_{i=1}^n\sum_{j=i+1}^n J_{ij}(\sigma_i,\sigma_j) - \sum_{i=1}^nh_i(\sigma_i).
\end{equation}
The probability distribution of this system is given by 
\begin{equation}
\pp(\vec \sigma) = \frac{e^{-\beta\HH(\vec \sigma)}}{\sum_{\vec \sigma'}e^{-\beta\HH(\vec \sigma')}}.
\end{equation}
This measure is invariant under the gauge transformation,
\begin{equation}\label{eq:gauge}
\begin{aligned}
h_{i}(\mu) &\to h_{i}(\mu) + \phi_i + \sum\nolimits_{j}^{j\ne i}\Lambda_{ij}(\mu)\\
J_{ij}(\mu,\nu) &\to J_{ij}(\mu,\nu) - \Lambda_{ij}(\mu) -\Lambda_{ji}(\nu) + \psi_{ij}
\end{aligned}    
\end{equation}
for any $\phi_i$, $\psi_{ij}$ and $\Lambda_{ij}(\mu)$. 
This gauge symmetry means that the Potts measure is characterised by ${\binom{n}{2}}(q-1)^2 + n (q-1)$ independent parameters, which is the same number of independent parameters in single and two-spin distributions, $\pp(\sigma_i)$ and $\pp(\sigma_i,\sigma_j)$ (see, e.g., Ref~\cite{Morcos:11}). 
Indeed for specified $\pp(\sigma_i)$ and $\pp(\sigma_i,\sigma_j)$ the Potts measure is the unique maximum-entropy model~\cite{Morcos:11}.
Another consequence of the gauge invariance is that a family of model parameters $(J,h)$ can result in the same measure.
As a result, inference methods that produce a unique set of parameters must invoke gauge fixing conditions (either explicitly or via implicit regularisations).
\subsection*{Mean-field inversion}
For completeness, we reproduce the derivation of the mean-field inversion method for Potts models from Ref~\cite{Morcos:11}.
We define the free energy 
\begin{equation}\label{eq:FF}
\FF = \FF(J,h) = -\ln \sum\nolimits_{\vec \sigma}e^{-\beta \HH(\vec \sigma)}
\end{equation}
It follows that the first and second-order derivatives of this free energy are related to the single-spin and pairwise distributions via
\begin{equation}\label{eq:dF/dh}
\frac{\del \FF}{\del h_{i\mu}} = -\pp_{i\mu}
\quad\text{and}\quad
\frac{\del^2 \FF}{\del h_{i\mu}\del h_{j\nu}} =
	-\pp_{i\mu,j\nu} + \pp_{i\mu}\pp_{j\nu}
\end{equation}
where we introduce the shorthand notations
\begin{equation}\nonumber
\begin{aligned}
h_{i\mu}		&=h_i(\sigma_i\!=\!\mu),	&
J_{i\mu,j\nu}	&=J_{ij}(\sigma_i\!=\!\mu,\sigma_j\!=\!\nu),
\\
\pp_{i\mu} 		&=\sum\nolimits_{\vec \sigma}\delta_{\sigma_i,\mu}\pp(\vec \sigma),	
&
\pp_{i\mu,j\nu} &=\sum\nolimits_{\vec \sigma}\delta_{\sigma_i,\mu}\delta_{\sigma_j,\nu}\pp(\vec \sigma).
\end{aligned}
\end{equation}
Eq~\eqref{eq:dF/dh} also implies 
\begin{equation}\label{eq:dP/dh}
\frac{\del\pp_{i\mu}}{\del h_{j\nu}} = \pp_{i\mu,j\nu} - \pp_{i\mu}\pp_{j\nu}\equiv C_{i\mu,j\nu}
\end{equation}
where $C_{i\mu,j\nu}$ denotes the connected correlation matrix.

\paragraph{Gauge fixing} To infer a unique set of model parameters, we adopt the lattice-gas gauge which explicitly limits the model parameters to those that are independent (see, Eq~\eqref{eq:gauge} and the text around it). 
In this gauge each spin has a gauge state, $c_i$ for spin $i$, for which the pairwise coupling and local field vanish, i.e.,
\begin{equation}\label{eq:lattice-gauge}
\forall \vec \sigma,i,j: J_{ij}(\sigma_i,c_j) = J_{ij}(c_i,\sigma_j) = h_i(c_i)=0
\end{equation}
We assume this gauge in the following analysis unless specified otherwise.

\paragraph{Legendre transformation}
Since the local field $h_{i\mu}$ is conjugate to the single-spin distributions $\pp_{i\mu}$ (see, Eq~\eqref{eq:dF/dh}), we can define a Legendre transform of the free energy 
\begin{equation}\label{eq:GG}
\GG =	\FF + \sum\nolimits_{i\mu} h_{i\mu}\pp_{i\mu}.
\end{equation}
Note that $\GG$ does not depends explicitly on the probability of the gauge state $\pp_{ic_i}$; it is left out of the summation by the gauge condition $h_{ic_i}\!=\!0$ [Eq~\eqref{eq:lattice-gauge}]. 
In this ensemble the local fields are given by
\begin{equation}\label{eq:local_fields}
h_{i\mu} = \frac{\del\GG}{\del \pp_{i\mu}}.
\end{equation}
Taking the derivative of the above equation yields
\begin{equation}\label{eq:inverse_connected_corr}
\frac{\del h_{i\mu}}{\del \pp_{j\nu}}
	= \frac{\del^2\GG}{\del \pp_{i\mu}\del \pp_{j\nu}} 
	= (C^{-1})_{i\mu,j\nu}
\end{equation}
where the last equality follows from Eq~\eqref{eq:dP/dh} and the fact that the first-order derivatives of a function and its Legendre transform are inverse functions of one another.
Note that the indices $(i\mu,j\nu)$ in Eqs~\eqref{eq:local_fields} and \eqref{eq:inverse_connected_corr} do not include the gauge states. 

\paragraph{Small-coupling expansion}
To derive the mean-field inversion, we consider a systematic expansion around the non-interacting Hamiltonian, treating the coupling term as a perturbation~\cite{Georges:91,Yedidia:01},
\begin{equation}\label{eq:HH_alpha}
-\beta\HH_\alpha(\vec\sigma) = \alpha\sum\nolimits_{i<j} J_{ij}(\sigma_i,\sigma_j) + \sum\nolimits_{i}h_i(\sigma_i)
\end{equation}
where the parameter $\alpha$ tunes the interaction strength: $\HH_0$ corresponds to the non-interacting case and $\HH_1$ to the original Hamiltonian. 
Expanding $\GG$ as a power series in $\alpha$ yields
\begin{equation}
\GG_\alpha =
\GG_{0} + \GG'_{0}\alpha + \tfrac{1}{2}\GG''_{0}\alpha^2 + \OO(\alpha^3)
\end{equation}
where $\GG'_{\alpha}=d\GG_{\alpha}/d\alpha$ and $\GG''_{\alpha}=d^2\GG_{\alpha}/d\alpha^2$.
Substituting the above expression in Eqs~\eqref{eq:local_fields} and \eqref{eq:inverse_connected_corr} gives
\begin{equation}
    \label{eq:samll_coupling_exp}
    \begin{aligned}
h_{i\mu}
&=
\frac{\del\GG_0}{\del \pp_{i\mu}}  
+ 
\frac{\del\GG'_0}{\del \pp_{i\mu}} \alpha 
+ 
\OO(\alpha^2)
\\
(C^{-1})_{i\mu,j\nu}
&=
\frac{\del\GG_0}{\del \pp_{i\mu}\del \pp_{j\nu}}  
+ 
\frac{\del\GG'_0}{\del \pp_{i\mu}\del \pp_{j\nu}} \alpha 
+ 
\OO(\alpha^2)
    \end{aligned},
\end{equation}
for $i\mu\ne ic_i$ and $j\nu \ne jc_j$. 

\paragraph{Zeroth order}
When $\alpha=0$, the spins decouple and the free energy reads
\begin{equation}
\FF_0 = -\sum\nolimits_i\ln \sum\nolimits_{\nu}e^{h_{i\nu}}
\end{equation}
From Eq~\eqref{eq:dF/dh}, we have $P_{i\mu}=e^{h_{i\mu}}/\sum_{\nu}e^{h_{i\nu}}$ and
\begin{equation}
\GG_0 
	=\sum_{i\mu\ne ic_i}\pp_{i\mu}\ln\pp_{i\mu}
	+
	\sum_i\left(1-\sum_{\nu\ne c_i}\pp_{i\nu}\right)
	\ln\left(1-\sum_{\nu\ne c_i}\pp_{i\nu}\right).
\end{equation}
Taking the derivatives, we have
\begin{equation}\label{eq:0order}
\frac{\del\GG_0}{\del\pp_{i\mu}} = \ln\frac{\pp_{i\mu}}{\pp_{ic_i}}
\quad
\text{and}
\quad
\frac{\del^2\GG_0}{\del\pp_{i\mu}\del\pp_{j\nu}} = \delta_{ij}\left(\frac{\delta_{\mu\nu}}{\pp_{i\mu}}+\frac{1}{\pp_{ic_i}}
\right),
\end{equation}
where $\pp_{ic_i}\!=\!1-\sum_{\mu\ne c_i}\pp_{i\mu}$.
We note that the pairwise coupling does not appear in the zeroth-order expansion. 

\paragraph{First order}
Differentiating the thermodynamic potential $\GG_\alpha$ with respect to $\alpha$ gives
\begin{equation}
\GG'_\alpha = 
	-\sum_{\vec\sigma}
	\frac{e^{-\beta\HH_\alpha(\vec\sigma)}}{\sum_{\vec \sigma'}e^{-\beta\HH_\alpha(\vec \sigma')}}
	\sum_{i<j} J_{ij}(\sigma_i,\sigma_j).
\end{equation}
Note that the expression for $\GG_\alpha$ can be obtained from Eqs~\eqref{eq:FF} and \eqref{eq:GG} for the small-coupling Hamiltonian in Eq~\eqref{eq:HH_alpha}. 
In the limit $\alpha\!\to\!0$, the Boltzmann weight becomes that of the non-interacting system and the above equation reduces to
\begin{equation}
\GG'_0 =
	-\sum\nolimits_{i<j} \sum\nolimits_{\mu\nu} \pp_{i\mu}\pp_{j\nu}J_{i\mu,j\nu}
\end{equation}
Therefore we have
\begin{equation}\label{eq:dGG1_dp}
\frac{\del\GG'_0}{\del\pp_{i\mu}} 
    = 
	-\sum\nolimits_{j\nu}^{j\ne i}\pp_{j\nu}J_{i\mu,j\nu}.
\end{equation}
Here the gauge condition on $J$ ensures that the single-spin probability of the gauge state does not appear on the \textit{r.h.s.}
Note that $J_{i\mu,j\nu}$ for $j<i$ does not enter the model and we let $J_{i\mu,j\nu}=J_{j\nu,i\mu}$ for convenience. 
Taking the derivative of Eq~\eqref{eq:dGG1_dp}, we obtain
\begin{equation}
\frac{\del^2\GG'_0}{\del\pp_{i\mu}\del\pp_{j\nu}} =  -(1-\delta_{ij})J_{i\mu,j\nu}
\end{equation}
Substituting Eq~\eqref{eq:0order} and the above equation in Eq~\eqref{eq:samll_coupling_exp} gives
\begin{equation}\label{eq:mf_inversion_J}
(C^{-1})_{i\mu,j\nu} \approx
	\left\{
	\begin{array}{ll}
	\frac{\delta_{\mu\nu}}{\pp_{i\mu}}+\frac{1}{\pp_{ic_i}}	&\text{if $i=j$}\\
	-\alpha J_{i\mu,j\nu}	&\text{if $j\ne i$}
	\end{array}
	\right.
\end{equation}
Finally we combine Eqs~\eqref{eq:samll_coupling_exp},\eqref{eq:0order} and \eqref{eq:dGG1_dp} to obtain the self-consistent condition for the local fields
\begin{equation}\label{eq:mf_inversion_h}
    h_{i\mu}=
    \ln\frac{\pp_{i\mu}}{\pp_{ic_i}}
    -
    \alpha \sum\nolimits_{j\nu}^{j\ne i}\pp_{j\nu}J_{i\mu,j\nu} 
    + 
    \OO(\alpha^2)
\end{equation}
The naive mean-field inversion method is based on Eqs~\eqref{eq:mf_inversion_J} and \eqref{eq:mf_inversion_h} which relate the model parameters to the empirically accessible connected correlation matrix. 

\subsection*{Phase transitions in Potts models on homogeneous random graphs}
Here we reproduce the mean-field analysis of Potts models (see, e.g., Ref~\cite[Sec.~IC]{Wu:82}).
Consider a uniform-interaction ferromagnetic $q$-state Potts model on a graph,
\begin{equation}\label{eq:true_H}
\HH(\vec{\sigma}) = -\sum_{(ij)\in\EE}\delta_{\sigma_i,\sigma_j},
\end{equation}
where $\delta_{\sigma_i,\sigma_j}$ denotes the Kronecker delta and the summation is over the graph's edges $\EE$. 
In the mean-field approximation, all spins are identical and the internal energy and entropy of the system read
\begin{equation}
\label{eq:U}
U = -|\EE|\sum\nolimits_{\mu=1}^q p_\mu^2
\quad\text{and}\quad
S = - n\sum\nolimits_{\mu=1}^q p_\mu \ln p_\mu
\end{equation}
where $p_\mu$ is the fraction of spins in state $\mu$, $n$ the number of spins and $|\EE|$ the numbers of edges (interactions).
To analyse the ferromagnetic transition, we consider the ansatz
\begin{equation}
\label{eq:Delta}
p_\mu = \frac{1}{q} (1 - \Delta) + \delta_{\mu,q}\Delta
\end{equation}
where $\Delta$ is the order parameter and we chose the state $q$ as the spin state of the ferromagnetic phase.
This ansatz yields the free energy per spin
\begin{equation}
    \begin{aligned}
    \beta(f(\Delta)-f(0)) ={}& 
    \frac{1+(q-1)\Delta}{q}\ln (1+(q-1)\Delta)
    \\&
    +
    \frac{q-1}{q}(1-\Delta)\ln(1-\Delta)
    -
    \frac{q-1}{2q}\frac{\lambda}{T}\Delta^2
    \end{aligned}
\end{equation}
where $\lambda\!=\!2|\EE|/n$ is the mean coordination number.
In the thermodynamic limit $n\!\to\!\infty$, a phase transition exists at the critical temperature 
\begin{equation}    \label{eq:Tc_MF}
    \frac{1}{T_c} =\frac{1}{\lambda}\times\left\{
    \begin{array}{ll}
    q&\text{if $q\le2$}\\
    2\frac{q-1}{q-2}\ln(q-1)&\text{if $q>2$}
    \end{array}
    \right.
\end{equation}
The free energy is minimised by $\Delta\!=\!0$ for $T\!>\!T_c$ and by the largest root of the equation
\begin{equation}
e^{-\lambda\Delta/T}=\frac{1-\Delta}{1+(q-1)\Delta}
\end{equation}
for $T\!<\!T_c$.
This phase transition is continuous for $q\!\le\!2$ and discontinuous for $q\!>\!2$ in which the order parameter and internal energy per spin are discontinuous across the transition, 
\begin{equation}
    \label{eq:potts_discontinuity}
    \begin{aligned}
    \Delta(T_c^-)-\Delta(T_c^+) &= \frac{q-2}{q-1}
    \\
    u(T_c^-)-u(T_c^+)&=-\lambda\frac{(q-2)^2}{2q(q-1)}.
    \end{aligned}
\end{equation}
Finally we note that the above analysis is exact for complete graphs in which all spins in the system are truly (as opposed to statistically) identical.

\begin{acknowledgments}
This work was supported in part by the Alfred P.\ Sloan Foundation, the Simons Foundation, the National Institutes of Health under award number R01EB026943 and the National Science Foundation, through the Center for the Physics of Biological Function (PHY-1734030). 
\end{acknowledgments}

%

\end{document}